\documentstyle[aps,preprint,epsf]{revtex}
\newcommand{\beq}{\begin{equation}}
\newcommand{\eeq}{\end{equation}}
\begin{document}
\draft
\tightenlines

\title{ Universal Behavior of Heavy-Fermion Metals Near a Quantum
Critical Point}

\author{ V.R.  Shaginyan$^{a,b}$ \footnote{E--mail:
vrshag@thd.pnpi.spb.ru}}
\address{$^{a}$Petersburg Nuclear Physics Institute,
Gatchina, 188300, Russia;\\
$^{b}$CTSPS, Clark Atlanta University, Atlanta,
Georgia 30314, USA}
\maketitle

\begin{abstract}

The behavior of the electronic system of heavy fermion metals
is considered. We show that there exist at least two main types of the
behavior when the system is nearby a quantum critical point which can be
identified as the fermion condensation quantum phase transition
(FCQPT). We show that the first type is represented by the behavior of
a highly correlated Fermi-liquid, while the second type is depicted
by the behavior of a strongly correlated Fermi-liquid.
If the system approaches FCQPT from the disordered phase,
it can be viewed as a highly correlated Fermi-liquid which at low
temperatures exhibits the behavior of Landau Fermi liquid (LFL).
At higher temperatures $T$, it demonstrates the non-Fermi liquid (NFL)
behavior which can be converted into the LFL behavior by the application
of magnetic fields $B$. If the system has undergone FCQPT,
it can be considered as a strongly
correlated Fermi-liquid which demonstrates the NFL behavior even at
low temperatures. It can be turned into LFL by applying magnetic
fields $B$.  We show that the effective mass $M^*$ diverges at the
very point that the N\'eel temperature goes to zero.  The $B-T$ phase
diagrams of both liquids  are studied. We demonstrate that these
$B-T$ phase diagrams have a strong impact on the main properties of
heavy-fermion metals such as the magnetoresistance, resistivity,
specific heat, magnetization, volume thermal expansion, etc.

\end{abstract}


In heavy-fermion (HF) metals with strong electron correlations,
quantum phase transitions at zero temperature may strongly influence
the measurable quantities up to relatively high temperatures.  These
quantum phase transitions have recently attracted much attention
because the behavior of HF metals
is expected to follow universal patterns defined by the quantum
mechanical nature of the fluctuations taking place at quantum
critical points (see e.g. \cite{sac,voj}).
It is widely believed that the proximity of the electronic
system of HF metal to quantum critical points
may lead to non-Fermi liquid (NFL) behavior.
The system can be driven to quantum critical
points (QCPs) by tuning control parameters
other then temperature, for example, by
pressure, by magnetic field, or by doping. When a system
is close to QCP we are dealing with the strong coupling limit where no
absolutely reliable answer can be given on pure theoretical first
principle grounds. Therefore, the only way to verify what type of
quantum phase transition occurs
is to consider experimental facts which describe the behavior of the
system.  Only recently, there appeared experimental facts which
deliver experimental grounds to understand the nature of quantum
phase transition producing the universal behavior of HF metals.

It is the very nature of HF metals that suggests
that their unusual properties
are defined by a quantum phase transition
related to the unlimited growth of the effective
mass at its QCP. Moreover, a divergence to
infinity of the effective electron mass
was observed  at a magnetic field-induced QCP \cite{geg,pag,cus}.
We assume that such a quantum phase transition to be
the fermion condensation quantum phase transition (FCQPT), an essential
feature of which is the divergence of the effective mass $M^*$
at its QCP \cite{ms,ks}.
FCQPT takes place when the density $x$ of system tends to the critical
density $x_{FC}$ so that $M^*\propto 1/r$
where $r$ is a distance from the QCP, $r=|x-x_{FC}|$.
Such a behavior does not qualitatively depend on the system's
dimensions and valid in both cases of two-dimensional (2D) and
three-dimensional (3D) Fermi systems \cite{shag1,khod}. Beyond FCQPT
the system possesses fermion condensation (FC) and represents a new
state of electron liquid with FC \cite{ks,vol}.  As soon as FCQPT
occurs, the system becomes divided into two quasiparticle subsystems:
the first is characterized by quasiparticles with the effective mass
$M^*_{FC}$, while the second one is occupied by quasiparticles with
mass $M^*_L$. The quasiparticle dispersion law in systems with FC
can be represented by two straight lines, characterized by the
effective masses $M^*_{FC}$ and $M^*_L$, and intersecting near the
binding energy $E_0$.  Properties of these new quasiparticles with
$M^*_{FC}$ are closely related to the state of the system which is
characterized by the temperature $T$, pressure $P$, or by the
presence of the superconductivity.  We may say that the quasiparticle
system in the range occupied by FC becomes very ``soft'' and is to be
considered as a strongly correlated liquid.  Nonetheless, the basis
of the Landau Fermi liquid theory \cite{lan} survives FCQPT: the low
energy excitations of the strongly correlated liquid with FC are
quasiparticles, while this state can be considered as a quantum
protectorate \cite{ms}.
The only difference between the Landau Fermi-liquid
and Fermi-liquid after FCQPT is that we have to expand the number of
relevant low energy degrees of freedom by introducing a new type of
quasiparticles with the effective mass $M^*_{FC}$ and the energy
scale $E_0$ \cite{ms}. It is possible to provide a consistent picture
of high-$T_c$ metals as the strongly correlated Fermi-liquid
\cite{ms1}.

When a Fermi system approaches FCQPT from
the disordered phase, its low energy excitations are
Landau quasiparticles
which can be characterized by the effective mass $M^*$.
This mass strongly depends on the distance $r$, temperature and
magnetic fields $B$ \cite{shag1}.
At low temperatures, it becomes a Landau Fermi liquid
with the effective mass $M^*$ provided that $r>0$.
This state of the system, with $M^*$ strongly depending on
$T$, $r$ and $B$, resembles the strongly correlated liquid.
In contrast to the strongly correlated liquid, there is no
energy scale $E_0$ and the system under consideration is
the Landau Fermi liquid at $T\to 0$. Therefore, this liquid
can be called a highly correlated liquid. Such a highly correlated
Fermi-liquid was observed in non-superconducting
La$_{1.7}$Sr$_{0.3}$CuO$_4$ \cite{shag1,nakam}.

In this Letter, we continue to show that within the framework of
FCQPT it is possible to understand the NFL behavior observed in
different strongly and highly correlated Fermi liquids such as
high-$T_c$ superconductors \cite{ms1} and heavy-fermion metals.
We apply the theory of fermion condensation
to describe the behavior of the electronic system of HF metals
and to show that there exist at least two main types of the
behavior. If the system approaches FCQPT from the disordered phase
it can be viewed as the highly correlated electron liquid, and
the effective mass $M^*$ depends on temperature, $M^*\propto
T^{-1/2}$.  Such a dependence of $M^*$ leads to the
NFL behavior of the electronic system.  The application
of a magnetic field $(B-B_{c})\geq B^*(T)\propto T^{3/4}$ restores
Landau Fermi liquid (LFL) behavior.  Here $B_{c}$ is a critical
magnetic field. At $(B-B_{c})\geq B^*(T)$, the effective mass depends
on the magnetic field, $M^*(B)\propto (B-B_{c})^{-2/3}$,  being
approximately independent of the temperature at $T\leq T^*(B)\propto
(B-B_{c})^{4/3}$. At $T\geq T^*(B)$, the $T^{-1/2}$ dependence of the
effective mass and the NFL behavior are re-established.
At $T\to 0$, the system becomes LFL
with the effective mass $M^*\propto 1/r$.
When the system has undergone FCQPT, it becomes a strongly
correlated electron liquid, and the effective
mass behaves as $M^*\propto 1/T$ leading to the NFL
behavior even at low temperatures.
The application of a magnetic field $(B-B_{c})\geq
B^*(T)\propto T^{2}$ restores the LFL behavior.  At
$(B-B_{c})\geq B^*(T)$, the effective
mass $M^*(B)\propto (B-B_{c})^{-1/2}$,  being
approximately independent of the temperature at $T\leq T^*(B)\propto
\sqrt{B-B_{c}}$.  At $T\geq T^*(B)$, both the $1/T$ dependence and
the NFL behavior are re-established.
We show that the effective mass $M^*$ diverges
at the very point that the N\'eel temperature goes to zero.
It is demonstrated
that obtained $B-T$ phase diagrams have a strong impact on the main
properties of HF metals such as the magnetoresistance, resistivity,
specific heat, magnetization, volume thermal expansion, etc.

We start with the case of a highly correlated electron liquid when
the system approaches FCQPT from the disordered phase.
FCQPT manifests itself in
the divergence of the quasiparticle effective mass $M^*$ as the
density $x$ tends to the critical density $x_{FC}$, or the distance
$r\to 0$ \cite{shag1,khod} \begin{equation} M^*\propto
\frac{1}{|x-x_{FC}|}\propto \frac{1}{r}.  \end{equation} Since the
effective mass $M^*$ is finite, the system exhibits the LFL behavior
at low temperatures $T\sim T^*(x)\propto|x-x_{FC}|^2$ \cite{shag1}.
The quasiparticle distribution function $n({\bf p},T)$ is given by
the equation \begin{equation} \frac{\delta \Omega}{\delta n({\bf
p},T)} =\varepsilon ({\bf p},T)-\mu (T)-T\ln \frac{1-n({\bf
p},T)}{n({\bf p},T)}=0.  \end{equation} The function $n({\bf p},T)$
depends on the momentum ${\bf p}$ and the temperature $T$.  Here
$\Omega=E-TS-\mu N$  is the thermodynamic potential, and $\mu $ is
the chemical potential, while $\varepsilon ({\bf p},T)$,
\begin{equation}
\varepsilon ({\bf p},T)=\frac{\delta E[n(p)]}{\delta n({\bf p},T)},
\end{equation}
is the quasiparticle energy. This energy is a functional of $n({\bf
p},T)$ just like the total energy $E[n(p)]$, entropy $S[n(p)]$ and
the other thermodynamic functions. The entropy $S[n(p)]$ is given by
the familiar expression \begin{equation} S[n(p)]=-2\int \left[ n({\bf
p},T)\ln n({\bf p},T)+(1-n({\bf p},T)) \ln (1-n({\bf p},T))\right]
\frac{d{\bf p}}{(2\pi )^3}, \end{equation} which results from purely
combinatorial considerations. Eq. (2) is usually presented as the
Fermi-Dirac distribution \begin{equation} n({\bf p},T)=\left\{ 1+\exp
\left[ \frac{(\varepsilon ({\bf p},T)-\mu )}T \right] \right\} ^{-1}.
\end{equation} At $T\to 0$, one gets from Eqs. (2), (5) the standard
solution $n_F({\bf p},T\to 0)\to \theta (p_F-p)$, with $\varepsilon
(p\simeq p_F)-\mu =p_F(p-p_F)/M^{*}$, where $p_F$ is the Fermi
momentum, $\theta (p_F-p)$ is the step function, and $M^{*}$ is the
Landau effective mass \cite{lan} \begin{equation} \frac
1{M^{*}}=\frac 1p\frac{d\varepsilon (p,T\to0)}{dp}|_{p=p_F}.
\end{equation}
It is implied that in the case of LFL $M^{*}$ is
positive and finite at the Fermi momentum $p_F$. As a result, the
$T$-dependent corrections to $M^{*}$, to the quasiparticle energy
$\varepsilon (p)$, and other quantities, start with $T^2$-terms
being approximately temperature independent.
The Landau equation relating the mass $M$ of an
electron to the effective mass of the quasiparticles is of the form
\cite{lan}
\begin{equation}
\frac{{\bf p}}{M^*}=\frac{{\bf p}}{M}+ \int F_L({\bf p},{\bf p}_1,x)
\nabla_{{\bf p}_1}n({\bf p_1}) \frac{d{\bf p}_1}{(2\pi)^3}.
\end{equation}
Applying Eq. (7) at $T<T^*(x)$, we obtain the common result
\begin{equation}
M^*=\frac{M}{1-N_0F^1_L(x)/3}.
\end{equation}
Here $N_0$ is the density of states of the free Fermi gas and
$F^1_L(x)$ is the $p$-wave component of the Landau interaction.
At $x\to x_{FC}$, the
denominator in Eq. (8) tends to zero and one obtains Eq. (1).
The temperature smoothing out the step function $\theta(p_F-p)$ at
$p_F=(x/3\pi^2)^{1/3}$ induces the variation of the
Fermi momentum $\Delta
p_F\sim TM^*/p_F$. We assume that the amplitude $F_L$
has a short range $q_0\ll p_F$ of interaction in the momentum space.
It is a common condition leading to the existence of FC
and nearly-localized Fermi liquids \cite{ks,pif}.
If the radius is such that $q_0\sim\Delta p_F\sim T_0M^*/p_F$,
corrections to the effective
mass are proportional to $T$ at $T\sim T_0$. Here
$T_0\propto |x-x_{FC}|$ is a characteristic temperature
at which the system's behavior is of the NFL type.
On the other hand, at $T^*(x)\ll T_0$,
we have $q_0\gg T^*(x)M^*/p_F$, and the system behaves as LFL  at
$T\sim T^*(x)$, so that the corrections to the effective mass start
with $T^2$ terms. We can also conclude that the transition region is
rather large compared with $T^*(x)$ being proportional to $T_0$.

In the case of $T\sim T_0$, we again can use Eq. (8) with
$F^1_L(p_F+\Delta p_F)\sim F^1_L(p_F)+A\Delta p_F$,
where $A\propto dF^1_L(x)/dx$.
Substituting this expansion of $F^1_L(p_F+\Delta p_F)$ into Eq. (8)
we find that \begin{equation} M^*\sim\frac{M}{A\Delta p_F}\propto
\frac{M}{M^*T}.  \end{equation} In deriving Eq. (9) we assumed that
the system is close to FCQPT so that $(1-N_0F^1_L(p_F))\ll N_0A\Delta
p_F$.  We can say that at $T\sim T_0$, $\Delta p_F$ induced by $T$
becomes larger then the distance $r$ from FCQPT, $\Delta
p_F>|p_F^{FC}-p_F|$, where $p_F^{FC}$ corresponds to $x_{FC}$.
Solving Eq. (9) with respect to $M^*$, we obtain \cite{shag1}
\begin{equation} M^*(T)\propto \frac{1}{\sqrt{T}}.  \end{equation}
The behavior of the effective mass given by Eq. (10) can be verified
by measuring the thermal expansion coefficient which is given by
\cite{lanl1} \begin{equation} \alpha (T)=\frac{1}{3}\left(
\frac{\partial (\log V)}{\partial T}\right) _P =-\frac x{3K}\left(
\frac{\partial (S/x)}{\partial x}\right) _T.  \end{equation} Here,
$P$ is the pressure and $V$ is the volume.  Substituting Eq. (4) into
Eq. (11), one obtains that in the LFL theory coefficient is of the
order $\alpha(T)\sim M^*_LT/p_F^2K$.  By employing Eq. (10), one
obtains that at $T\sim T_0$ \cite{alp} \beq\alpha(T)\propto
a\sqrt{T}+bT,\eeq with $a$ and $b$ being constants.  This result is
in good agreement with experimental facts obtained in measurements on
CeNi$_2$Ge$_2$ \cite{geg1}.

The application of magnetic fields $B$ leads to  Zeeman
splitting of the Fermi level. As a result,
two quasiparticle distribution
functions with Fermi momenta
$p_{F}^1$ and $p_{F}^2$ appear, so that
$p_F^1<p_F<p_F^2$ and $\Delta p_F=(p_F^2-p_F^1)\sim \mu_0BM^*/p_F$.
Here $\mu_0$ is the electron magnetic moment.
In the same way Eq. (10) was derived, we can obtain the equation
determining $M^*(B)$ \cite{shag1}. The only difference is that there
are no contributions coming from the terms proportional to $\Delta
p_F$, and we have to take into account terms proportional to $(\Delta
p_F)^2$.  Assuming that the system is close to the critical point, we
obtain \beq M^*(B)\sim M\left(\frac{\varepsilon_F}
{B\mu}\right)^{2/3}. \eeq
At $T\sim T^*(x)$, Eq. (13) is valid as long as
$M^*(B)\leq M^*(x)$, otherwise we have to use Eq. (1).
It follows from Eq. (13) that the application of magnetic fields
reduces the effective mass.  If there exists a
magnetic order in the system which is suppressed by magnetic field
$B=B_{c0}$, then the quantity $(B-B_{c0})$ plays
the role of zero field, and
Eq. (13) has to be replaced by the equation,
\beq M^*(B)\propto \left(\frac{1}
{B-B_{c0}}\right)^{2/3}.\eeq
At high magnetic fields, we expect Eq. (14) to be invalid
because $\Delta p_F$ becomes too large so that $\Delta p_F>q_0$.
In that case, the effective mass still depends on the magnetic
field  but the proportionality given by Eq. (14) is not preserved,
and the dependence on the magnetic field becomes weaker.

At elevated temperatures $T\sim T_0$, the effective mass starts to
depend on both the temperature and the magnetic field. A cross over
from the $B$-dependent effective mass $M^*(B)$ to the $T$-dependent
effective mass $M^*(T)$ takes place at a transition temperature
$T^*(B)$ as soon as $M^*(B)\simeq M^*(T)$. This requirement and Eqs.
(10) and (14) give that \beq T^*(B)\propto (B-B_{c0})^{4/3}. \eeq At
$T>T^*(x)$, Eq. (15) determines the line on the $B-T$ phase diagram
which separates the region of the LFL behavior taking place at
$T<T^*(B)$ from the NFL behavior occurring at $T>T^*(B)$. At
$T<T^*(B)$, the system behaves like LFL with the effective mass
$M^*(B)$, and corrections to the effective mass start with $T^2$
terms. In accordance with the LFL theory, the specific heat $c\simeq
\gamma T$, with \beq \gamma(B)\propto M^*(B)\propto
(B-B_{c0})^{-2/3}.\eeq The resistivity $\rho$ behaves as
$\rho=\rho_0+A(B)T^2$, where the coefficient \beq A(B)\propto
(M^*(B))^2\propto (B-B_{c0})^{-4/3}.\eeq It follows from Eq. (16) and
(17) that the Kadowaki-Woods ratio $K=A/\gamma^2$ \cite{kadw} is
conserved. All these results obtained from Eqs.  (14-17) are in good
agreement with experimental facts observed in measurements on the HF
metal YbAgGe single crystal \cite{bud}.  The critical behavior of the
coefficient $A(B)\propto (B-B_{c0})^{\beta}$ at $B\to B_{c0}$
described by Eq. (14) with $\beta =-4/3$ is in accordance with
experimental data obtained in measurements on CeCoIn$_5$  which
displayed the critical behavior with $\beta =-1.37\pm 0.1$
\cite{pag}.

In the LFL theory, the magnetic susceptibility
$\chi\propto M^*/(1-F^a_0)$. Note, that
there is no ferromagnetic instability in Fermi systems related
to the growth of the effective mass, and the
relevant Landau amplitude $F^a_0>-1$ \cite{pif}.
Therefore, at $T<T^*(B)$, the magnetic susceptibility
turns out to be proportional to the effective mass
\beq \chi(B)\propto M^*(B)\propto (B-B_{c0})^{-2/3},\eeq
while the static magnetization $M_B(B)$ is given by
\beq M_B(B)\propto B\chi\propto (B-B_{c0})^{1/3}.\eeq
At $T>T^*(B)$, as it follows from Eq. (10), Eq. (18) has to be
rewritten as \beq \chi(T)\propto M^*(T)\propto
\frac{1}{\sqrt{T}}.\eeq The behavior of $\chi(B)$ and $M_B(B)$ as a
function of magnetic field $B$ given by Eqs. (18) and (19) and the
behavior of $\chi(T)$, see Eq. (20), are in accordance with facts
observed in measurements on CeRu$_2$Si$_2$ with the critical field
$B_{c0}\to 0$ \cite{tak}.

Consider the system when $r\to 0$. Then its properties
are determined by the magnetic fields $B$ and the temperature $T$
because there are no other parameters to describe the state of the
system.  At the transition temperatures $T\simeq T^*(B)$, the
effective mass depends on both $T$ and $B$, while at $T\ll T^*(B)$,
the system is LFL with the effective mass being given by Eq. (14),
and at $T\gg T^*(B)$, the mass is defined by Eq. (10).  Instead of
solving Eq. (8), it is possible to construct a simple interpolation
formula to describe the behavior of the effective mass over all the
region, \beq M^*(B,T)\propto F(B,T)= \frac{1}{c_1(B-B_{c0})^{2/3}+
c_2f(y)\sqrt{T}}.\eeq Here, $f(y)$ is a universal monotonic function
of $y=\sqrt{T}/(B-B_{c0})^{2/3}$ such that $f(y\sim 1)=1$, and
$f(y\ll 1)=0$.  It is seen from Eq. (21) that the behavior of the
effective mass can be represented by a universal function $F_M$ of
only one variable $y$ if the temperature  is measured in the units of
the transition temperature $T^*(B)$, see Eq. (15), and the effective
mass is measured in the units of $M^*(B)$ given by Eq. (14),
$F_M(y)=aF(B,T)/M^*(B)$ with $a$ being a constant. This
representation describes the scaling behavior of the effective mass.
As seen from Eqs. (19) and (21), the scaling behavior of the
magnetization can be represented
in the same way,
provided the magnetization is normalized
by the saturated value at each field given
by Eq. (19)
\beq \frac{M_B(B,T)}{M_B(B)}\propto\frac{1}{1+ c_3f(y)y},\eeq
where $c_3$ is a constant.
It is seen from Eq. (22), that
magnetization is a monotonic function of $y$.
Upon using the definition of susceptibility, $\chi=\partial
M_B/\partial B$, and differentiating both sides of Eq. (22) with
respect to $B$, we arrive at the conclusion that the susceptibility
also exhibits the scaling behavior and can be presented as a
universal function of only one variable $y$, provided it is
normalized by the saturated value at each field given by Eq. (18)
\beq \frac{\chi(B,T)}{\chi(B)}\propto\frac{1}{1+ c_3f(y)y}
+2c_3y\frac{f(y)+ydf(y)/dy}{(1+ c_3f(y)y)^2}.\eeq
It is of importance
to note that the susceptibility is not a monotonic function of $y$
because the derivative is the sum of two contributions.  The second
contribution on the right hand side of Eq. (23) makes the
susceptibility have a maximum. The above behaviors of the
magnetization  and susceptibility are in accordance with the facts
observed in measurements on CeRu$_2$Si$_2$ \cite{tak}.
Note, that the magnetic properties of CeRu$_2$Si$_2$ do not show
any indications of the magnetic ordering
at the smallest temperatures and in the smallest
applied magnetic fields \cite{tak}, that is $B_{c0}\to 0$ in that case.
As a result, we can conclude that the QCP is driven by the divergence of
the effective mass rather then by magnetic fluctuations, and
FCQPT is the main cause of the NFL behavior.  We can also
conclude that the N\'eel temperature is zero in this case,
because the magnetic susceptibility diverges at $T\to0$, as
it is seen from Eq. (20). A more
detailed analysis of this issue will be published elsewhere.

Consider the case when the system has undergone FCQPT.
Then, there exist special solutions of Eq. (2)
associated with the so-called fermion condensation \cite{ks}. Being
continuous and satisfying the
inequality $0<n_0({\bf p})<1$ within some
region in $p$, such solutions $n_0({\bf p})$
admit a finite limit for the logarithm in Eq.
(2) at $T\rightarrow 0$ yielding  \cite{ks}
\begin{equation}
\varepsilon({\bf p})-\mu=0,\quad \mbox{if}
\quad 0<n_0({\bf p})<1;\,p_i\leq p\leq p_f,
\end{equation}
where  $\varepsilon({\bf p})$ is given by Eq. (3).
At $T=0$, Eq. (24) defines a new state of electron
liquid with FC \cite{ks,vol},
which is characterized by a flat spectrum in the
$(p_f-p_i)$ region, and which can strongly
influence measurable quantities up to temperatures
$T\ll T_f$.
In this state, the order parameter of the superconducting state
$\kappa ({\bf p})=\sqrt{(1-n_0({\bf p}))n_0({\bf p})}$
has finite values in the $(p_f-p_i)$ region, whereas the
superconducting gap $\Delta _1\to 0$ in this region, provided that
the pairing interaction tends to zero.
Such a state can be considered as superconducting, with an
infinitely small value of $\Delta_1$, so that the entropy $S(T=0)$
of this state is equal to zero  \cite{ms,ks}.

When $p_f\to p_i\to p_F$ the flat part vanishes, and Eq. (24)
determines QCP at which the effective mass $M^*$ diverges and FCQPT
takes place. When the density
approaches QCP from the disordered phase, Eq. (24) possesses
non-trivial solutions at $x=x_{FC}$ as soon as the effective
inter-electron interaction as a function of the density,
or the Landau amplitude,
becomes sufficiently strong to determine the
occupation numbers $n({\bf p})$ which delivers the minimum value to
the energy $E[n(p)]$, while the kinetic energy can be considered as
frustrated \cite{ms}.
As a result, the occupation numbers $n({\bf p})$ become
variational parameters and Eq. (24) has non-trivial solutions
$n_0({\bf p})$, because the energy $E[n(p)]$ can be lowered by
alteration of the occupation numbers. Thus, within the region
$p_i<p<p_f$, the solution $n_0({\bf p})$ deviates from the Fermi step
function $n_F({\bf p})$ in such a way that the energy
$\varepsilon({\bf p})$ stays constant,
while outside this region $n({\bf p})$ coincides with
$n_F({\bf p})$ \cite{ks}.  Note, that a formation of the flat part of
the spectrum has been confirmed in Ref. \cite{dzy,lid,irk}.

At $r>0$ when the system is on the disordered side, that is
$\kappa ({\bf p})\equiv 0$, and the density $x$
moves away from QCP located at $x_{FC}$, the Landau
amplitude $F_L(p=p_F,p_1=p_F,x)$ as a function of $x$
becomes smaller, the kinetic energy comes into a play  and
makes the flat part vanish. Obviously, Eq. (24) has only the trivial
solution $\varepsilon (p=p_F)=\mu $,
and the quasiparticle occupation
numbers are given by the step function, $n_F({\bf p})=\theta
(p_F-p)$.

At $\Delta_1\to 0$, the critical temperature $T_c\to 0$.
We see that the ordered phase can exist only at $T=0$,
and the state of electron liquid with FC disappears at $T>0$
\cite{ms}.  Therefore, FCQPT is not the endpoint of a line of
finite-temperature phase transitions. This conclusion is in
accordance with Eq. (2) which does not admit the existence of the
flat part of spectrum at finite temperatures.
As a result, the quantum to classical crossover upon approaching
a finite-temperature phase transition does not exist. In the
considered case, one can expect to observe such a cross-over at
$T\sim T_f$. On the other hand,
$\Delta_1$ becomes finite if we assume that the pairing
interaction is finite, and the
corresponding $x-T$ phase diagram becomes
richer. Moving along this line, we can consider the high-$T_c$
superconductivity as well, see e.g. \cite{ms,ks}.

At finite temperatures $T\ll T_f$, the occupation numbers in the
region $(p_f-p_i)$ are still determined by Eq. (24), and the system
becomes divided into two quasiparticle subsystems:  the first
subsystem is occupied by normal quasiparticles with the finite
effective mass $M_L^{*}$ independent of $T$ at momenta $p<p_i$, while
the second subsystem in the $(p_f-p_i)$ range is characterized by the
quasiparticles with the effective mass $M_{FC}^{*}(T)$
\cite{ms,khod1} \begin{equation} M_{FC}^{*}\simeq
p_F\frac{p_f-p_i}{4T}.  \end{equation} There is an energy scale $E_0$
separating the slow dispersing low energy part, related to the
effective mass $M_{FC}^{*}$, from the faster dispersing relatively
high energy part, defined by the effective mass $M_L^{*}$. It follows
from Eq. (25) that $E_0$ is of the form \cite{ms}
\begin{equation}E_0\simeq 4T.\end{equation}

The described system can be viewed as a strongly correlated one,
it has the second type of the behavior and
demonstrates the NFL behavior even at
low temperatures.
By applying magnetic fields, the system can be driven to LFL with
the effective mass \cite{shag}
\beq M^*(B)\propto \frac{1}{\sqrt{B-B_{c0}}}.\eeq
In the same way as it was done above, we find
from Eqs. (25) and (27) that a cross over from the
$B$-dependent effective mass $M^*(B)$
to the $T$-dependent effective mass
$M^*(T)$ takes place at a transition temperature $T^*(B)$
\beq T^*(B)\propto \sqrt{(B-B_{c0})}. \eeq
Equation (28) determines the line in the $B-T$ phase
diagram which separates the region of
the LFL behavior at $T<T^*(B)$ from
the NFL behavior occurring at $T>T^*(B)$.
The existence of the $B-T$ phase diagram given by
Eqs. (15) and (28) can be highlighted by
calculating the resistivity and the magnetoresistance \cite{shag1}.
The resistivity, which at $T>T^*(B)$ demonstrates the NFL behavior,
at $T<T^*(B)$,  exhibits the LFL behavior,
$\rho=\rho_0+A(B)T^2$.
The $B-T$ diagram  of the dependence of the effective mass on the
magnetic field can be highlighted by
calculating the magnetoresistance.
At $(B-B_{c0})>B^*(T)$, the magnetoresistance  is
negative, and at $(B-B_{c0})<B^*(T)$, it becomes positive.
This behavior of both the magnetoresistance and the resistivity
is in agreement with measurements on YbRh$_2$Si$_2$ \cite{geg},
when the system exhibits the second type
of the behavior, see Eq. (28),
while CeCoIn$_5$ and YbAgGe demonstrate the first type of behavior
consistent with that given by Eq. (15) \cite{pag,bud}.

At $T<T^*(B)$, the coefficients $\gamma\propto M^*(B)$,
$\chi(B)\propto M^*(B)$,
and $A(B)\propto (M^*(B))^2$, and we find that the Kadowaki ratio
$K$ and the Sommerfeld-Wilson ratio $R\propto \chi(B)/\gamma(B)$
are preserved due to Eq. (27). The obtained
$B-T$ phase diagram and the conservation of both the Kadowaki and
the Sommerfeld-Wilson ratios are in full agreement with data obtained
in measurements on YbRh$_2$Si$_2$
and YbRh$_2($Si$_{0.95}$Ge$_{0.05}$)$_2$ \cite{geg,cus,geg1}.
Taking into account Eqs. (11) and (25), we obtain that
in the case of the two quasiparticle subsystems
the thermal expansion coefficient
$\alpha(T)\propto a+bT+c\sqrt{T}$,
with $a$, $b$ and $c$ being constants.
Here, the first term $a$
is determined by the FC contribution, the second $bT$ is given
by normal quasiparticles with the effective mass
$M_L^{*}$, and
the third $c\sqrt{T}$ comes from a specific
contribution related to the spectrum $\varepsilon_c({\bf p})$ which
insures the connection between the dispersionless region $(p_f-p_i)$
occupied by FC and normal quasiparticles \cite{alp,khod1}.
At finite temperatures, the contribution coming from the third term
is expected to be relatively small because the spectrum
$\varepsilon_c({\bf p})$ occupies a relatively small area in the
momentum space.
Since at $T\to0$, the main contribution to the specific heat $c(T)$
comes from the spectrum $\varepsilon_c({\bf p})$, the specific heat
behaves as  $c(T)\propto a_1\sqrt{T}+b_1T$, with $a_1$ and $b_1$
being constants. The second term $b_1T$ comes from the contribution
given by FC and normal quasiparticles. Measurements for
YbRh$_2($Si$_{0.95}$Ge$_{0.05}$)$_2$ show a power low divergence of
$\gamma=c/T\propto T^{-\alpha}$ with $\alpha=1/3$ \cite{geg1}.
This result is in a reasonable agreement with our calculations giving
$\alpha=0.5$. At lower temperatures, the relative contribution of the
first term $a_1\sqrt{T}$ becomes bigger, and we expect that the
agreement will also become better.  Now we obtain that the Gr\"uneisen
ratio ${\rm \Gamma}(T)=\alpha(T)/c(T)$ diverges as ${\rm
\Gamma}(T)\propto 1/\sqrt{T}$ \cite{alp}. This results is in good
agreement with measurements on YbRh$_2($Si$_{0.95}$Ge$_{0.05}$)$_2$
\cite{cus,geg1}.

As it follows from Eq. (27),
the static magnetization behaves as $M_B(B)\propto \sqrt{B-B_{c0}}$
in accordance with measurements on
YbRh$_2($Si$_{0.95}$Ge$_{0.05}$)$_2$
\cite{cus}. We can also conclude that
Eqs. (21), (22), and (23) determining
the scaling behavior of the effective mass, static magnetization and
the susceptibility are also valid in the case of
strongly correlated liquid, but the variable $y$ is now given by
$y=T/\sqrt{B-B_{c0}}$, while the function $f(y)$ can be dependent
on $(p_f-p_i)/p_F$.  This dependence comes from Eq. (25).
As a result, we can obtain that at $T<T^*(B)$, the factor
$d\rho/dT\propto A(B)T$ behaves as $A(B)T\propto T/(B-B_{c0})$, and
at $T>T^*(B)$, it behaves as $A(B)T\propto 1/T$.  These observations
are in good agreement with the data obtained in measurements on
YbRh$_2($Si$_{0.95}$Ge$_{0.05}$)$_2$ \cite{cus}.

It is worthy to note that in zero magnetic fields the N\'eel
temperature is zero at FCQPT, because, as it follows from Eq.
(10), the effective mass tends to infinity at FCQPT and makes the
susceptibility be divergent. On the other hand,
if there is the magnetic order and the N\'eel temperature
is not equal to zero, the effective mass is
finite and there is no FCQPT.
As soon as the magnetic order is suppressed at
$B\to B_{c0}$, that is the N\'eel temperature
tends to zero, the effective mass
$M^*(B)\to \infty$ as it follows from Eq. (14).
If the system has undergone FCQPT, again at $B\to B_{c0}$,
the N\'eel temperature goes to zero
and $M^*(B)\to \infty$, see Eq. (27). If $B_{c0}=0$,
the effective mass diverges at $T\to 0$, see Eq. (25),
and the susceptibility $\chi$ tends to infinity being proportional to
the effective mass. In this case, the N\'eel temperature is equal
to zero as well. Therefore, one may say that
the effective mass $M^*$ diverges at the very point that the N\'eel
temperature goes to zero.

A few remarks are in order at this point.
To describe the behavior of heavy-fermion metals,
we have introduced the system of quasiparticles, as it is done in the
Landau theory of normal Fermi-liquids,
where the existence of fermionic quasiparticles is a generic
property of normal Fermi systems independent of microscopic details.
As we have seen, at $T\ll T_f$, these quasiparticles have universal
properties which determine the universal behavior of heavy-fermion
metals.  One can use another approach constructing the singular part
of the free energy, introducing notions of the upper critical
dimension, hyperscaling, etc., see e.g. \cite{sac,voj}.
Moving along this way, one may expect difficulties. For example,
having the only singular part, one has to describe at least the two
types of the behavior. We reserve a consideration of these items for
future publications.

In conclusion, we have shown that our simple
model based on FCQPT explains the
critical behavior
observed in different HF metals.
In the case of such HF metals as
CeNi$_2$Ge$_2$, CeCoIn$_5$, YbAgGe, CeRu$_2$Si$_2$, etc., the behavior
can be explained by the proximity to
FCQPT, where their electronic systems
behave like highly correlated liquids.
In the case of such HF metals as YbRh$_2($Si$_{0.95}$Ge$_{0.05}$)$_2$
and YbRh$_2$Si$_2$ the critical behavior
is different. This can be explained by
the presence of FC in the electronic
systems of these metals, i.e., by the fact
that the electronic systems have undergone FCQPT and behave
as strongly correlated liquids.
We have shown that the basis of the Landau Fermi liquid
theory survives in the both cases: the low energy
excitations of both strongly correlated Fermi-liquid with FC and
the highly correlated Fermi-liquid are
quasiparticles. It is also shown that
the effective mass $M^*$ diverges at the very point that the N\'eel
temperature goes to zero.
The $B-T$ phase diagrams of both the highly correlated liquid and
the strongly correlated one have been studied. We have shown
that these $B-T$ phase diagrams influence strongly
the effective mass and such important properties of HF metals
as  magnetoresistance, resistivity, specific heat,
magnetization, susceptibility, volume thermal expansion, etc.

I thank P. Coleman for valuable comments.
I am grateful to the CTSPS for hospitality during
my stay in Atlanta. This work was supported in part by the Russian
Foundation for Basic Research, No 04-02-16136.

\end{document}